\documentclass[aps,prl,twocolumn,floats,showpacs]{revtex4}
  \usepackage{epsfig}
  \usepackage{amsmath}
  \usepackage{amssymb}
\begin{document}
\title{Quantum Dot in the Kondo Regime coupled to p-wave superconductors\\}
\author{Tomosuke Aono $^{1}$, Anatoly Golub $^1$ and Yshai Avishai
$^{1,2}$ } 
\affiliation {$^1$Department of Physics and $^2$Ilse Katz Center,
Ben-Gurion University, Beer-Sheva, Israel }
\begin{abstract}
This paper studies the physics of junctions containing 
superconducting $(S)$ and normal $(N)$ leads 
weakly coupled to an Anderson impurity  in the Kondo regime $(K)$. 
Special attention is devoted to the case where one of the 
leads is a $p-wave$ superconductor where 
mid-gap surface states play an important role in the tunneling
processes and help the formation of Kondo resonance.
The novel physics in these systems 
beyond that encountered in quantum dots 
coupled only to to normal leads is 
that electron transport at finite bias $eV$ 
in $SKN$ and $SKS$ junctions is governed by 
Andreev reflections. These enable the occurrence of 
dissipative current even when the bias $eV$ is smaller 
than the superconducting gap $\Delta$. 
Using the slave boson mean field approximation 
the current,
shot-noise power and Fano factor are calculated
as functions of the applied
bias voltage in the sub-gap region $eV < \Delta$
and found to be strongly
dependent on the ratio $t_K$ between the Kondo temperature $T_{K}$ and
the superconducting gap $\Delta$. In particular, for large values 
of $t_K$ the attenuation of current due to the existence of the
superconducting gap is compensated by the Kondo effect. 
This scenario is manifested also in the behavior of the Josephson 
current as function of temperature. 
\end{abstract}
\maketitle
\section{introduction}
The Kondo effect is one of the simplest manifestations 
of many-body physics exhibiting 
strong correlations\cite{hewson}.
Originally, it has been introduced 
in order to explain the occurrence of a shallow 
minimum of the
 resistivity $\rho(T)$ of bulk metals doped with a 
small concentration of
magnetic impurities. It has been confirmed that at 
low temperatures 
(below the Kondo temperature $T_K$) 
scattering from magnetic impurities is strong, 
leading to an enhanced resistivity. It is now 
well established that at this 
strong coupling regime the theory of perturbation fails due 
to logarithmic divergences of higher (than second) order diagrams. 

The Kondo physics also  plays
an important role in electron
transport through quantum dots, where 
instead of a magnetic impurity one now encounters localized electrons
 \cite{glazman88,hershfield91,ournca}. 
Inspection of the dot-electron Green function 
reveals an enhancement
 of the local density of states at the Fermi energy 
(the Abrikosov-Suhl resonance). Thus, unlike the case of
Kondo effect in bulk metals, its hallmark 
experimental manifestation in quantum dots is an 
enhancement of the {\it conductance}. An important milestone
in this field has been recorded by
recent observations of the Kondo effect in transport
through quantum dots\cite{goldhaber,reviews} and in a finite
length carbon nanotubes \cite{lin}. 
These experiments have paved the way for a new
class of experimental investigations 
into the physics of strongly correlated
electrons in general and the Kondo effect in particular. 
Together with more recent experiments
\cite{unitary}
demonstrate the feasibility of exploiting tunable physical
parameters of a quantum dot system in order to yield important
information on Kondo systems 
and other many-body related phenomena. 
Such studies,  for example, encompass 
 the entire crossover region between the Kondo 
limit, the mixed valence
regime and the non-Kondo (weak-coupling) domain.

While the Kondo physics in a quantum dot attached on both its 
sides to normal ($N$) metallic leads received much recent attention, 
it has been realized that
novel physical effects emerge if (one or both) electrodes attached 
through a tunneling barrier to 
a quantum dot in the Kondo regime ($K$) is a superconductor ($S$).
The central electron-transport mechanism 
in such $SKS$ and $SKN$ junctions is that of
Andreev reflections when two particles tunnel together coherently
to form a Cooper pair in the superconductor.
In fact, the physics of $SKS$ and $SKN$ junctions 
is determined by the interplay of these Andreev
reflections with the
Kondo-formed resonance in the spectral density of 
states of the dot electron \cite{BTK}.
This interplay between the Kondo effect and
superconductivity in the sub-gap region is 
rather effective here. 
We then expect the 
the $I-V$ characteristics in the sub-gap region
to be most important. 
Andreev reflections play a dominant role both in and out of equilibrium.
In the former case they are responsible for occurrence of direct
Josephson current while in the latter case they are the cause of
dissipative currents at sub-gap voltages
as well as for suppressing the zero bias anomaly in $NS$ junctions.
The effect becomes even more significant when the effective Kondo
temperature $T_{K}$ exceeds the superconducting energy gap $\Delta$. 
A natural candidate for such $SKN$ junction is that in which 
the role of the quantum dot or the Kondo impurity is played 
by a Carbon nano-tube $(CNT)$, where 
relatively high values of $T_{K}$ can be achieved.
Indeed,
 Kondo effect with rather high Kondo temperature $(T_{K} > 1K$)
has been reported in $N-CNT-N$ junctions
with a few conducting channels \cite{lin}. Moreover,
superconducting junctions with a weak link formed by $CNT$ 
have already been fabricated
\cite{Kas,Christ}. 

Let us illuminate the somewhat subtle distinction between 
$s-wave$ and $p-wave$ $SKN$ junctions.
In the case of an $SKN$
junction when one electrode is an $s-wave$ superconductor and the
Kondo impurity is weakly coupled to the $S$ and 
$N$ electrodes, is determined by
the competition between two important effects. The first one is
the formation of a Kondo singlet which screens the bare 
impurity spin and drives the system toward the unitary 
limit at very low temperatures. The second one is the existence
of the superconducting gap which implies 
a vanishingly small density of low
energy electron states
\cite{fazio,raimondi,ambe}. These are precisely the electron 
states which are needed in order to screen the Kondo impurity. 
Note that if the Kondo temperature is larger than the 
superconducting gap ($T_{K}>\Delta$) then electron states 
outside the gap can participate in the screening interaction. 
However, this condition cannot be easily realized. 
Consider, on the other hand, an $SKN$ junction in which 
the $S$ electrode 
consists of an {\it unconventional} superconductor which is
oriented relative to the interface in such a way that the pair potential
reverses its sign on the Fermi surface. In this case,
mid-gap zero-energy states $(ZES)$ are formed which are
localized near the surface of the unconventional
superconductor. These states can now participate 
in screening the impurity spin through the
Kondo effect and emergence of sub-gap current is expected. 
In the experimentally feasible setup of $S-CNT-N$ where 
just a few tunneling channels are present,
charge is carried mainly by quasiparticles moving
perpendicular to the interface. This restricts the 
possible values of the angle $\theta$ between the 
superconducting surface and the direction of
the injected quasiparticles \cite{tanaka}.
Formation of $ZES$ is possible when
$\Delta(\theta)=-\Delta(\pi-\theta)$. 
If the impurity is almost point-like, the relevant injection 
angle is of course $\theta=0$.
Unconventional
superconductors for which the symmetry of the pair potential is
that of a triplet ($p-wave$ superconductors) satisfy the above 
condition. These have recently been discovered by Maeno, {\it et al}
\cite{maeno} in $Sr_{2}RuO_{4}$. 
For impurities of finite extent, one may also consider 
 formation of $ZES$ in $d-wave$ superconductors.
Thus, the physics of 
$SKN$ junctions with an $S$ electrode whose order 
parameter has a non-trivial symmetry is affected by the  
formation of $ZES$ in the Kondo regime.

Beyond investigating the conductance dependence on the applied
bias we propose and explore theoretically other novel experimental
tools suitable for probing the Kondo regime. These include 
shot-noise measurements and the Josephson (direct) current.
Like the conductance, we will 
calculate and analyze shot-noise power spectrum in
$SKN$ junctions at very low temperature where one lead is either 
an $s-wave$ superconductor 
(mid-gap surface states are absent) or 
a $p-wave$ superconductor (mid-gap surface states are present). 
We will also study the temperature dependence of the Josephson 
current in these junctions. 

In order
to study this strongly interacting non-equilibrium problem in the
framework of workable approximation, we restrict ourselves to
voltages that do not exceed the superconducting gap so that the
Kondo temperature can satisfy the inequality 
$T_{K}> max (\Delta,eV)$. Under this
 condition we can apply the well known 
approach familiar in the physics of the Anderson impurity
Hamiltonian \cite{bickers}, namely 
the  mean field slave boson approximation
(MFSBA) (see below). It consists of a workable scheme 
for calculating the conductance and the
zero frequency shot-noise power. (Here we also apply the MFSBA to
calculate Josephson current in $SKS$ junctions.)

The combined effect of multiple Andreev reflections
(MAR) and electron-electron interactions on
the shot-noise in $SKS$ junctions was studied in Ref.
\onlinecite{us1,compagnano,arovas2}. 
It was shown that in the strong coupling Kondo
regime the effective transmission is enhanced, and a ballistic-like
channel opens up. Hence, an interplay between MAR
and the Kondo resonance yields an excess current 
at zero bias and the $I-V$
curve behaves similarly to that
of noninteracting ballistic
junctions. At very high values of $T_K$ and in the low voltage
limit the current approaches the noninteracting value
$I_{AR}=4e\Delta /h$. Analogously, the shot-noise power is shown
to display a pronounced maximum at $V=0$ and a decay as $1/V$ at
small bias,familiar in the standard theory of noninteracting $SNS$
junctions \cite{averin}.

In section II the Hamiltonian of $SKN$ and $SKS$ junctions is defined, 
and the bare action is introduced. After integrating out the 
fermion fields pertaining to the superconducting leads an 
effective action is obtained, depending solely on the dot variables. 
Calculations and presentations of conductance, shot-noise power 
and Josephson current are respectively detailed in sections 
III, IV and V.
\section
{Model Hamiltonian and effective action}
The dynamics of systems like $SKN$ or $SKS$ 
junctions is governed by the
Hamiltonian
\begin{eqnarray}
&& H=H_{L}+H_{R}+H_{d}+H_{t}+H_{c}, \label{H}
\end{eqnarray}
in which $H_{j}$, ($j=L,R$) are the  Hamiltonians of the electrodes which
depend on the electron field operators $\psi_{a\sigma}({\bf
r},t)$ where ${\bf r}=(x,y)$ and $\sigma=\pm$ is the spin index.
\begin{eqnarray}
\bbox{H}_j=\int dr [\Psi^{\dagger}_{j \sigma}
(\bbox{r})\xi({\bf \nabla})\Psi_{j}
        -\gamma \Psi^{\dagger}_{j 
\uparrow}(\bbox{r})\Psi^{\dagger}_{j \downarrow}
        \Psi_{j \downarrow}(\bbox{r})\Psi_{j \uparrow}(\bbox{r})].
\label{BCS}
\end{eqnarray}
Here $\gamma$ is the BCS coupling constant and
$\xi ({\bf \nabla})=-{\bf \nabla}^2/2m-\mu$ with $\mu$ being the 
chemical potential at temperature $T$. 
The Planck constant is set $\hbar =1$ and whenever appropriate,
the spin, space and time dependence of all the field operators 
will not be explicitly displayed.
As in Refs. \onlinecite{us1,us} the dot is 
represented by a single
level Anderson impurity with energy $\epsilon_{0}<0$ and Hubbard
repulsion parameter $U$. In the Kondo regime of interest here we
set $U \to \infty$ and assume $|\epsilon_{0}|$ to exceed any other
energy scale except $U$. In this case it is convenient to express
the dot and the tunneling Hamiltonians $H_{d}$ and $H_{t}$ via
slave boson (operators $b,b^{\dagger}$) and slave fermion
(operators $c,c^{\dagger}$) auxiliary fields\cite{coleman}.
Explicitly, 
\begin{eqnarray}
&& H_{d}=\epsilon_{0} \sum_{\sigma}c_{\sigma}^{\dagger}
c_{\sigma}, \nonumber \\
&& H_{t}=\sum_{j\sigma}{\cal T}_{j}
 c^{\dagger}_{\sigma} b\psi_{j \sigma}({\bf
0},t)+h.c.,
\label{Hdt}
\end{eqnarray}
 where ${\cal T}_{j}$ is the tunneling amplitude.
Finally, the Hamiltonian of the system must also include a term
which prevents double occupancy in the limit $U\rightarrow\infty$.
This term reads,
\begin{eqnarray}
&& H_{c}=\lambda(\sum_{\sigma}c^{\dagger}_
{\sigma}c_{\sigma}+b^{\dagger}b-1),
\label{Hc}
\end{eqnarray}
where $\lambda$ is a Lagrange multiplier.

Following Ref. \cite{us1} let us consider the dynamical ``partition
function''
\begin{eqnarray}
&& Z \sim \int {\cal D}[F] \exp(i{\cal S}),
\end{eqnarray}
where the path integral is carried out over all fields $[F]$ and
the action ${\cal S}$ is obtained by integrating the Lagrangian
pertaining to the Hamiltonian (\ref{H}) along the Keldysh contour.
In performing the functional integrations the boson field
operators are treated as $c$-numbers.  
As a result one arrives at an effective
action expressed in terms of the Green functions of the leads.
\begin{eqnarray}
&& S_{\rm eff}=-i{\rm Tr} \ln\hat{G}^{-1}- \int
dt[\hat{\lambda}\sigma_{z}(\hat{b}\hat{b}-1)]. \label{Seff2}
\end{eqnarray}
Here $\hat{\lambda}=(\lambda_{1},\lambda_{2}) $,
$\hat{b}=(b_{1},b_{2})$ and $\sigma_{z}$ are diagonal matrices
acting in Keldysh space. The
inverse propagator $\hat{G}^{-1}$ depends on the Green functions
of the electrodes.\onlinecite{us}. 
Performing the standard basis rotation in
Keldysh space one finds,
\begin{eqnarray}
&&\hat{G}^{-1}(\epsilon,\epsilon')=\delta
(\epsilon-\epsilon')(\epsilon -\tau_z\tilde{\epsilon}-
\frac{\Gamma b^2}{2} \tau_{z}\hat {g}_{+}(\epsilon)\tau_z),
\label{Eq_Ginv}
\end{eqnarray}
where $\tilde{\epsilon} = \epsilon_{0}+\lambda$ is the
renormalized level position (in the Kondo limit one has
$\tilde{\epsilon}\simeq 0$) and $\Gamma=(\Gamma_{L}+\Gamma_{R})/2
\propto {\cal T_{L,R}}^{2}$ is the usual transparency parameter.
The $2 \times 2$ matrix representation (in Keldysh space) for $g$ 
is composed of
diagonal elements $\hat {g}^{R/A}(\epsilon)$ and an
 upper off-diagonal element
$\hat {g}^{K}(\epsilon)=(\hat {g}^{R}-\hat {g}^{A})th(\epsilon/2T)$.
Here and below we define
\begin{eqnarray}
&& \hat {g}_{\pm} = \gamma_{L}\hat {g}_{L}\pm\gamma_{R}\hat
{g}_{R}, 
\label{hatg}
\end{eqnarray}
with  asymmetry
parameters $\gamma_{j}=\Gamma_{j}/\Gamma$.
 The matrix $\hat{g}_{R}$ has the standard structure
with retarded and advanced  superconductor Green functions which
in the $s-wave$ case reads,
\begin{eqnarray}
&& \hat {g}^{R/A}(\epsilon )= i\frac{(\epsilon \pm i0) +|\Delta|\tau_x}
{\sqrt{(\epsilon \pm i0)^2-|\Delta|^2}}, \label{Eq_g0RA}
\end{eqnarray}
while for the $p-wave$ case with incidence angle $\alpha$ it reads,
\begin{eqnarray}
&& \hat {g}^{R/A}(\epsilon)=\hat {g}_
{1}^{R/A}+\hat {g}_{2}^{R/A}\tau_{x}, \nonumber \\
\mbox{with} \\
&& \hat {g}_{1}^{R/A}= \frac{i\sqrt{(\epsilon\pm0)^2 - |\Delta|^2} \cos
\alpha - \epsilon \sin \alpha} {\epsilon\cos \alpha
+i\sqrt{(\epsilon\pm0)^2
-|\Delta|^2} \sin \alpha}\nonumber\\
&& \hat {g}_{2}^{R/A}= \frac{i|\Delta|}{\epsilon\cos \alpha  +i
\sqrt{(\epsilon\pm0)^2 - |\Delta|^2} \sin \alpha}.
\label{gpwave}
\end{eqnarray}
Note that the Pauli matrices $\tau_{x,y,z}$ act
in Nambu space.
For a one channel system one
can take $\alpha $=0 and simplify the expression for the Green
functions,
\begin{eqnarray}
&& \hat {g}_{1}^{R/A}= \frac{i\sqrt{(\epsilon\pm 0)^2 - 
|\Delta|^2}} {\epsilon\pm 0}\nonumber\\
&& \hat {g}_{2}^{R/A}= \frac{i|\Delta|}{\epsilon\pm 0 }.
\label{g12}
\end{eqnarray}
In the case of $NKS$ junction, 
the left lead is represented by the Keldysh Green's function of a
normal metal $\hat {g}_{L}$: $\hat {g}_{L}^{A/R}=\pm i  $\;; the Keldysh Green
function $\hat {g}_{L}^{K}$ has only diagonal matrix elements equal to
$2ith[(\epsilon\pm eV)/2T]$.
 Performing the
variation of the effective action with respect to the fields $b$ and
$\lambda$  a couple of self-consistency equations are 
obtained that
determine these fields. In order to explicitly write down these
self-consistency equations let us introduce the bare Kondo
temperature $ T_{K}^{0}= D exp[-\pi\|\epsilon_{0}|/(2\Gamma)]$ and
define a parameter $X$ by
$ \Gamma b^{2}=T_{K}^{0}X$, where $D$ is the energy
bandwidth. Then the MFSBA equations take the form
\begin{eqnarray}
&&X =-\frac{i\Gamma}{2T_K^0}{\rm Tr}\hat{G}^{K}\tau_{z}, 
\label{Eq x}\\
&& \lambda = \frac{i\Gamma}{8}{\rm
Tr}[\hat{G}^{K}\tau_{z}(\hat {g}_{+}^{R}+\hat {g}_{+}^{A})+
(\hat{G}^{R}+\hat{G}^{A})\tau_{z} \hat {g}_{+}^{K}]\tau_{z},
 \label{Eq y}
\end{eqnarray}
where the trace includes energy integration as well. Eq. (\ref{Eq x})
effectively determines the Kondo temperature (through the
parameter $X$), and reflects the constraint which prevents double
occupancy in the limit $U\rightarrow \infty$. The second
self-consistency equation (\ref{Eq y}) defines the renormalized
energy level position $\tilde{\epsilon}$. Let us briefly discuss
the validity range of the present analysis. The MFSBA is known to
encode the Kondo Fermi-liquid behavior at low temperatures. An
important parameter here is the ratio between the Kondo
temperature and the superconducting gap $t_K \equiv
T_{K}^{0}/\Delta$ \cite{fazio,raimondi,ambe}. For $t_K\gtrsim 1$ a
Fermi liquid behavior is expected. Accordingly, in this regime Eq.
(\ref{Eq x}) should have a nonzero solution $X\neq 0$ which
corresponds to nonzero $T_K$. On the other hand, in the limit of
large $\Delta$ the only possible solution is the trivial one $b=0$
(and, hence, $T_{K}=0$).  Quantitatively, the MFSBA
is reliable only for sufficiently large values of $t_K$. We
believe, however, that it can provide useful qualitative
information also for moderate values of $t_K$ describing a
crossover between the Kondo regime and the Coulomb blockade
dominant domain \cite{us}. 
It is worth noting here that, strictly speaking, the applied bias
voltage $V$ attenuates the Kondo resonance and lowers
$T_{K}$. Hence, for the reliability of the MFSBA in
non-equilibrium situations, both $\Delta$ and $eV$ should not
exceed the Kondo temperature. Attention below is mainly focused on
the sub-gap voltage regime $eV \lesssim \Delta$ in which case $t_K$
appears to be the only relevant parameter.
\section{Conductance}
 The expression for the tunneling current obtained in Ref.
\cite{us1} has a simple representation in terms of the dot 
Green function,
\begin{eqnarray} 
&&I=i\frac{eXt_K}{8\hbar}{\rm Tr}[(\hat{G}^{R}\tau_{z}-
\tau_{z}\hat{G}^{A})\hat {g}^{K}_{-} -\hat{G}^{K}\tilde{g}], 
\label{I}
\end{eqnarray}
where for $SKN$ junctions we denote
\begin{eqnarray}
&& \tilde{g} =-\gamma_{R}(\hat{g}^{R}\tau_{z}-\tau_{z}\hat{g}^{A})
-2i\gamma_{L}\tau_{z}, \nonumber\\
&&\tilde{g}^{R,A}=-\gamma_{R}(\tau_{z}\hat{g}^{R,A}+
\hat{g}^{R,A}\tau_{z}). 
\label{tildg}
\end{eqnarray}
The last quantity will be used later on as it
enters the expression for the noise power spectrum.
 Being combined with eqs.
(\ref{Eq x}) and (\ref{Eq y}) the result (\ref{I}) can be
conveniently used for computing the transport current and the
differential conductance of an $SKN$ junction in the Kondo regime
for different values of $t_K$.

For sufficiently large $t_K$ we anticipate a strong Kondo
resonance and the $G-V$ curves for both $s-wave $ and
$p-wave$ superconductors is expected to resemble that of purely
ballistic junctions  (for  $\gamma_{R}=\gamma_{L}=1$ ) without
interaction\cite{BTK}. Indeed, in the limit of large $t_K\gg 1$
which corresponds to the unitary, pure ballistic case, our
expression for the current (\ref{I}) reduces to that derived in Ref.
\cite{BTK}. This agreement is further supported by our numerical
calculations carried out for $t_K=100$ (see Fig.1). Calculations of
the current and conductance were also performed for
$t_K=5$,$3$, and $2$. The results are displayed in figure
1. For voltages in the sub-gap region, 
the conductance of an S(p-wave)KN junction  is distinct 
from that of an S(s-wave)KN junction. Such 
difference shows that the ZES supports the formation of a
Kondo singlet for the lower values $t_K=5$, $3$ and $2$ 
and effectively turn the junction to be
 more transparent, approaching  the BTK limit \cite{BTK}. 
 \vskip  -1.0 truecm
\begin{center}
\leavevmode \epsfxsize=3.5in \epsfbox{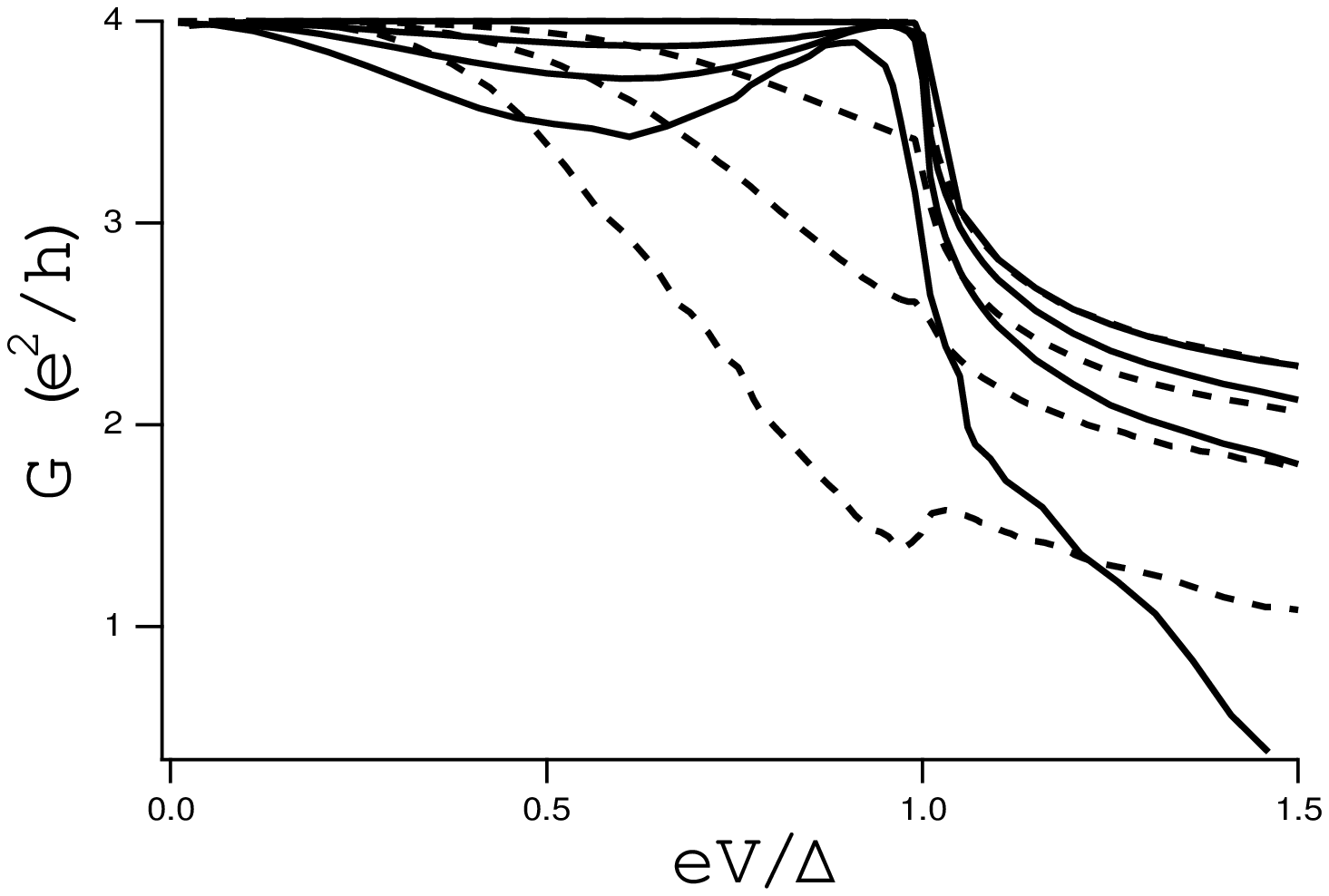}
\end{center}
\begin{small}
{\bf Fig. 1} The conductance $G$ (in units of $e^{2}/h$)
versus the bias $V$ (in units of $\Delta /e$) for an $
s-wave$ SKN (dash curves) and $S p-wave KN$ (solid lines) junction
at sub-gap voltages with $\Gamma/T_{K}^{0}=200$. The parameter
$t_{K}$ =2,3,5,100 (from down to top). The upper line 
corresponding to $t_{K}$
=100 coincides for s and p wave superconducting leads.
\end{small}
\vskip 1.5 truecm

Let us briefly summarize our results for the current and
conductance of SKN  junctions for $ s$ and $p$ wave
superconductors. In the limit of large $t_K$ the $G-V$ curve
is practically independent of $t_K$ and resembles that of a
ballistic junction, as indicated in the upper curve in Fig.1.
 For
lower values of $t_K$, junctions with s-wave superconductor lead
are driven away from the unitary limit and the underlying physics
becomes much richer. It reflects the influence of both $\Delta$
and $V$ on the Kondo resonance and on the actual value of the
Kondo temperature.
For  $t_K=5$ the $G-V$ curve noticeably deviates from that
obtained in the non-interacting limit. For $t_K=2$ the competition
between gap-related suppression of the Kondo effect and the
effective transparency of the junction becomes essential, leading
to further decrease of the conductance. However, it is interesting
to note that for and S(s-wave)KN junction at 
 $t_K=2$ the conductance displays a small peak 
  at the gap edge. Its interpretation is that the Kondo
correlations  strongly compete with superconductivity 
and influence 
the quasiparticle correlations 
when the energy exceeds the gap.
Such
 an effect takes place even when $t_{K}<1$ (see \cite{ambe}).
On the other hand, the S(p-wave)KN
 junction is less influenced to variations of 
 $t_K$. This may be explained by the fact that due to 
the presence of ZES,
 superconductivity plays a minor role in the formation of the Kondo
 resonance and BTK behavior persists for smaller values of $t_{K}$.
\section{shot-noise}

 The shot-noise spectrum is usually defined as
the symmetrized current-current correlation function
\cite{lesovik,rev}
\begin{eqnarray}
K(t_{1},t_{2})&=&\hbar [\langle I(t_{1})I(t_{2})\rangle
 - \langle I\rangle^2], \label{Eq_K12}
\end{eqnarray}
where  $\langle ... \rangle$ denotes quantum averaging with the
Hamiltonian (\ref{H}). The general expression for the zero frequency
shot-noise power in junctions with one 
(or even two) superconducting
leads (in MFSBA approximation) was obtained in our recent work
\cite{us1}. Like in the case of SKS junctions, it is useful to
write $K=(K_{1}+K_{2})e^{2}\Delta/(8\hbar)$ for which 
the expressions derived are,
\begin{eqnarray}
K_{1}= \frac{Xt_K}{2} {\rm Tr} \{
(\hat{g}^{R}_{+}-\hat{g}^{A}_{+})
(\hat{G}^{R}-\hat{G}^{A})-\hat{g}_{+}^{K}\hat{G}^{K} \},
\label{Eq_K1}
\end{eqnarray}
\begin{eqnarray}&&
K_{2}=-\frac{(Xt_K)^2}{8}{\rm Tr} \{(\hat{G}^{K}\tilde{g})^{2}
-2\tau_{z}\tilde{g}\tau_{z} \hat{G}^{A}\tilde{g}\hat{G}^{R}-
\nonumber \\
&& [2\tilde{g}\hat{G}^{R}
\tau_{z}\hat{g}^{K}_{-}\hat{G}^{K}+(\tilde{g}^{R}\hat{G}^{R})^{2}-
(\hat{G}^{A}\hat{g}^{K}_{-}\tau_{z})^{2}+{\rm h.c.}] \}.
\label{Eq_K2}
\end{eqnarray}

Expressions (\ref{Eq_K1}) and (\ref{Eq_K2}) (supplemented by the
self-consistency eqs. (\ref{Eq x}) and (\ref{Eq y})) are then solved
numerically for the same set of parameters $\Gamma/T_{K}^{0}=200$,
$t_{K}=100$, 5, 3 and $2$. The results for the shot-noise power
spectrum $K$ versus the applied voltage $V$ are displayed in figure
2. These results are clearly correlated with those for the
$G-V$ curve and can be summarized as follows:
In the limit $t_K \gg 1$ the characteristics of shot-noise 
power spectrum for both $s$ and $p$ wave
superconductors are consistent with those obtained for purely
ballistic junctions which exhibit strong suppressing of
the shot-noise power in the sub-gap region. At lower $t_K$ the physics
is distinct. For $t_K=5$ the noise spectrum for $s$ wave
superconducting lead still shows features typical for a junction
with relatively high transparency, while the results for $t_K=2$
are more similar to those for a low transparency junction. Such 
dependence is explicitly exposed in the plot of the Fano factor
versus the applied voltage (see Fig.3). Though the Fano factor does
not reach the maximum value of 2, it is strongly 
enhanced for the smaller
value of $t_K=2$. For $p-wave$ superconductor the shot-noise power (as
function of voltage) reflects the same physics as in the
conductance: ZES makes the Kondo resonance less vulnerable to the
impact of superconductivity and the junction
 remains close to the unitary
limit nearly in the whole range of values of $t_K$
considered here (see
the solid curves on Fig.2 and Fig.3).
\begin{center}
\leavevmode \epsfxsize=3.5in \epsfbox{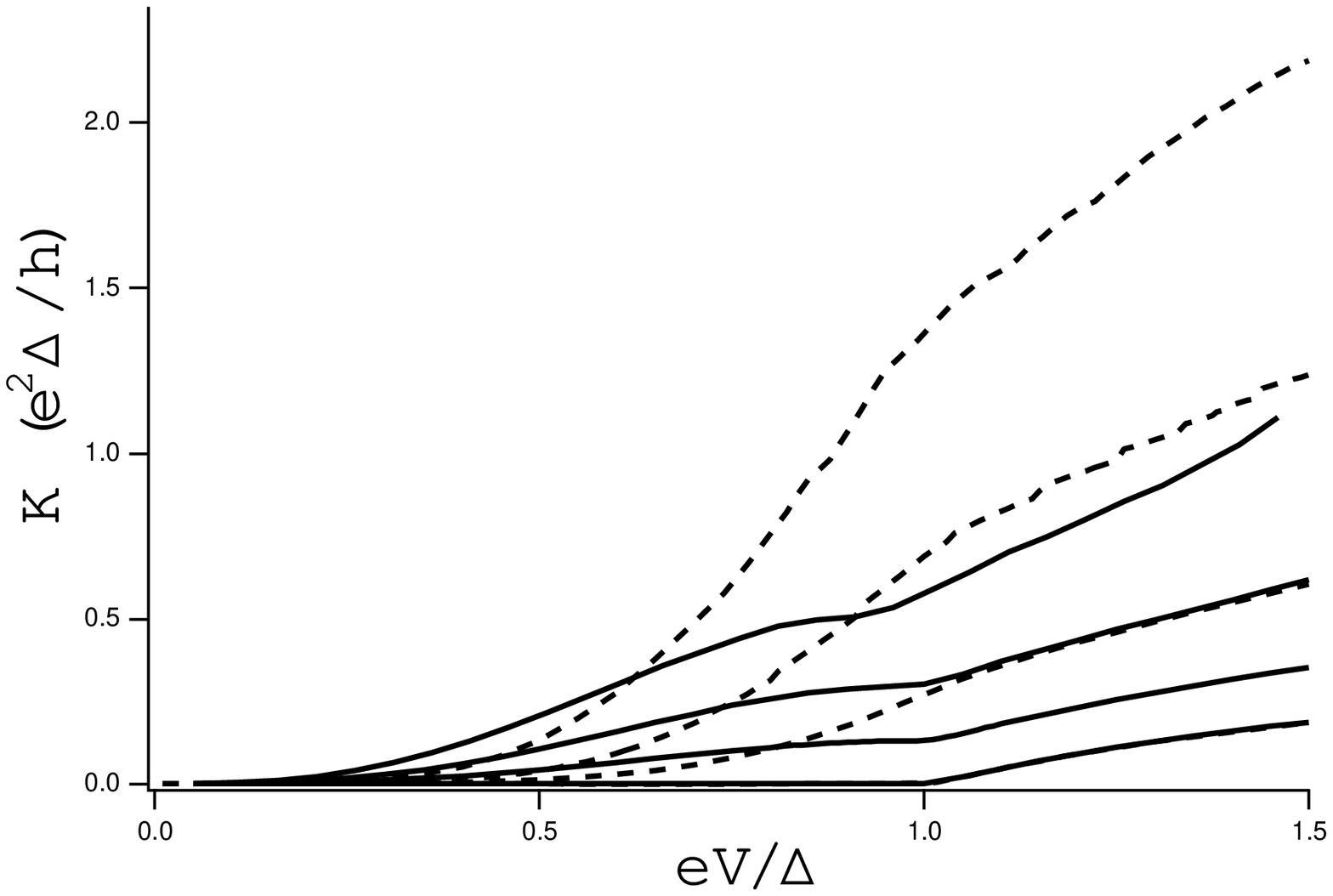}
\end{center}
\begin{small}
{\bf Fig. 2} The shot-noise power $K$ (in units of $e^{2}
\Delta/h$)  as a function of $V$ (in units of $\Delta /e$) for an
$SKN$ junction. The parameters and notations are the same as in
Fig. 1.($t_{K}$ grows from top to down. The curves with $t_{K}$=100
coincide for both s and p wave superconductors.\\
\end{small}
\vskip 1.5 truecm
\begin{center}
\leavevmode \epsfxsize=3.5in \epsfbox{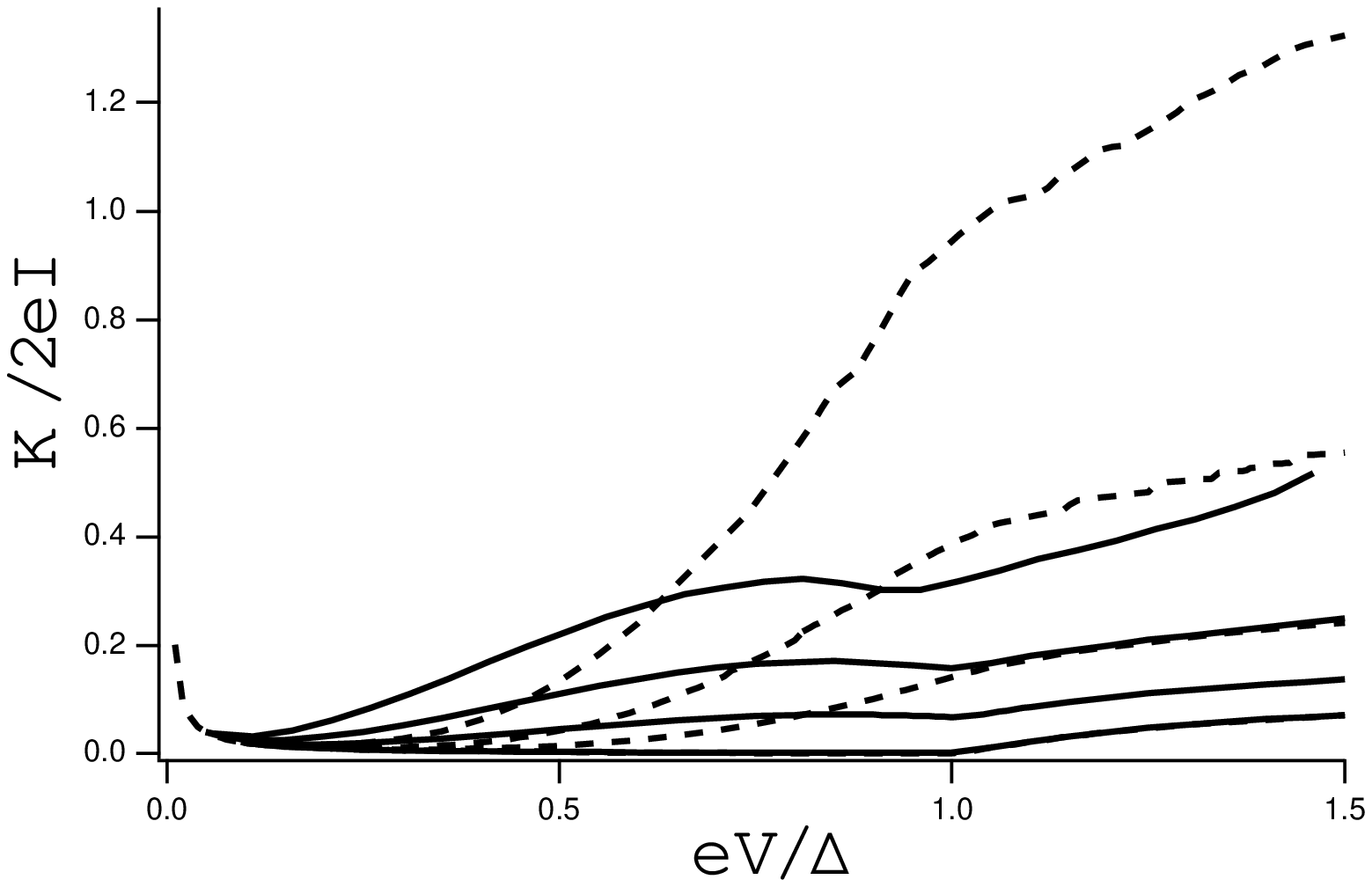}
\end{center}
\begin{small}
{\bf Fig. 3} The Fano factor $K/2eI$ 
as a function of $V$ (in units of $\Delta /e$) for
an $SKN$ junction.
The parameters and notations are the same as in Fig. 1.\\
\end{small}
\section{Josephson current}
In this section we study an
equilibrium property (Josephson effect) of an SKS junction in which
both electrodes are $p-wave$  superconductors (for comparison we
also represent the results for s-wave superconducting leads).
In the Kondo regime, the self-consistent equations for the $p-wave$
case(\ref{Eq x}) and (\ref{Eq y}) now read
\begin{eqnarray}
&& \widetilde{\epsilon} + \frac{2\Gamma}{\pi} \log
\frac{\widetilde{\epsilon}}{T_K}= \nonumber \\
&& \sum_{\omega}
\Bigl[
2 \Gamma
\frac{(1+ \alpha(\omega))\sqrt{\omega^2+\Delta^2}+ \beta(\omega) \Delta^2
/\omega \cos^2
\frac{\delta}{2}}{(1+\alpha(\omega))^2 \omega^2 + \widetilde{\epsilon}^2
+ \beta(\omega)^2 \Delta^2 \cos^2 \frac{\delta}{2}} 
\nonumber \\
&&
-\frac{2 \Gamma |\omega|}{\omega^2+ \widetilde{\epsilon}^2}
\Bigr ] 
\end{eqnarray}
\begin{eqnarray}
\widetilde{\Gamma}&=& \Gamma\sum \left[
 \frac{2 \widetilde{\epsilon}}{(1+\alpha(\omega))^2 \omega^2 +
\widetilde{\epsilon}^2 + \beta(\omega)^2 \Delta^2 \cos^2
\frac{\delta}{2}} \right] \label{Eq_yy}
\end{eqnarray}

Here $ \alpha(\omega) =   \widetilde{\Gamma}
\sqrt{\omega^2+\Delta^2}/\omega^2 $, $ \beta(\omega) =
\widetilde{\Gamma}/\omega $ and $\delta$ is the phase difference
between two superconductors. 
Self consistent equations for the case of 
$s-wave$ superconducting leads were derived in Ref. \cite{arovas2} 

The Josephson current is obtained by
variation of free energy with respect to the phase 
of the order parameter,
\begin{equation} 
I = \frac{2 e}{\hbar} \frac{\partial
F}{\partial \delta}. 
\label{JCdef}
\end{equation}
Explicitly we then find,
\begin{eqnarray}
I = \frac{e}{\hbar} \sum_{\omega} \frac{(\beta(\omega) \Delta)^2
\sin \delta } {\omega^2 (1+\alpha(\omega))^2 +
\widetilde{\epsilon}^2 + (\beta(\omega) \Delta)^2 \cos^2
\frac{\delta}{2}}
 \label{Eq_j}
\end{eqnarray}
The self-consistency equation and the expression for the 
Josephson current can easily be extended to the case of
an anisotropic coupling :$\Gamma_{L/R}=\Gamma(1\pm
p)$. 
For this, one should replace
$ \cos^2 \frac{\delta}{2} \rightarrow (\cos^2 \frac{\delta}{2}
+p^2 \sin^2 \frac{\delta}{2}) $ and
 $
I \propto  \sin \delta \rightarrow I \propto (1-p^2) \sin \delta$.

For a small anisotropy parameter $p=0.1$ and 
for values of $t_{K}$
=$100,5,3,2$, we calculate the Josephson current
 as a function of the phase
difference $\delta$ at temperatures close to $T=0$. The plot is
depicted in Fig.4. 
\begin{center}
\leavevmode \epsfxsize=3.5in \epsfbox{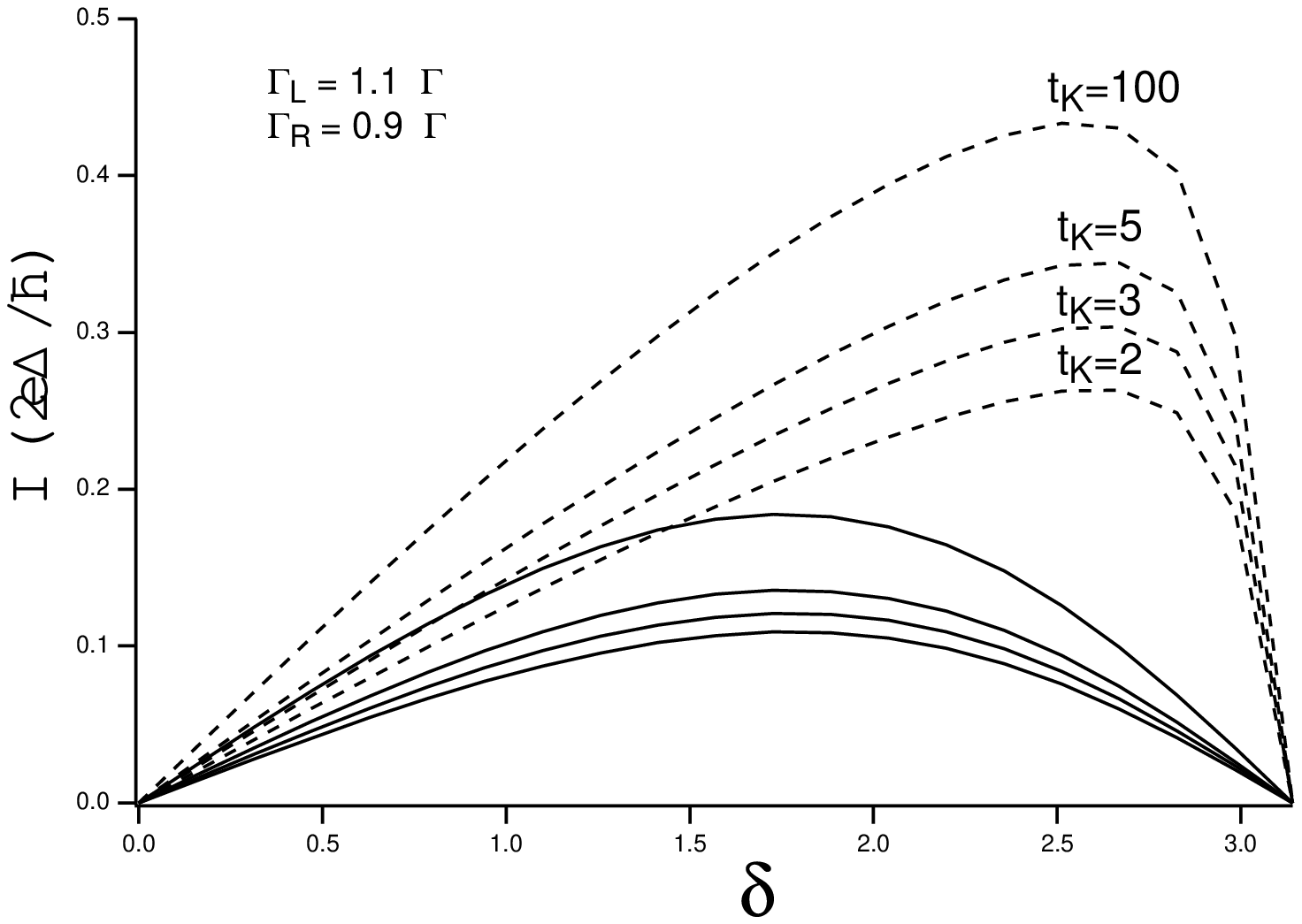}
\end{center}
\begin{small}
{\bf Fig. 4} Josephson current versus phase difference $\delta$ at
$T\rightarrow 0$. Dash and solid curves correspond to s and p-wave
superconductors, correspondingly.The parameters are the same as in
Fig.1. The  value of $t_{K}$ decreases  from top downward.
 \\
\end{small}
\vskip -0.5 truecm
In the $s-wave$ case (dashed lines in
figure 4) the current at $t_{K}$ =$100 $ qualitatively corresponds
to the unitary limit \cite{glazman}, while its 
amplitude is decreased
at smaller values of $t_{K}$. Nevertheless, in this region of parameters,
the Kondo effect strongly competes with
superconductivity and therefore there are no traces of a
$\pi$-junction. The mid-gap states for $p-wave$ superconductors
act as if they increase the effective normal region of
the junction and therefore, 
the weight of the contribution to the Josephson current 
which is proportional to
$sin\delta$ is reduced.
When $t_{K}$ becomes smaller,
the Josephson current for $p-wave$ junction tends
to be similar to that in an SIS junction.
\begin{center}
\leavevmode \epsfxsize=3.5in \epsfbox{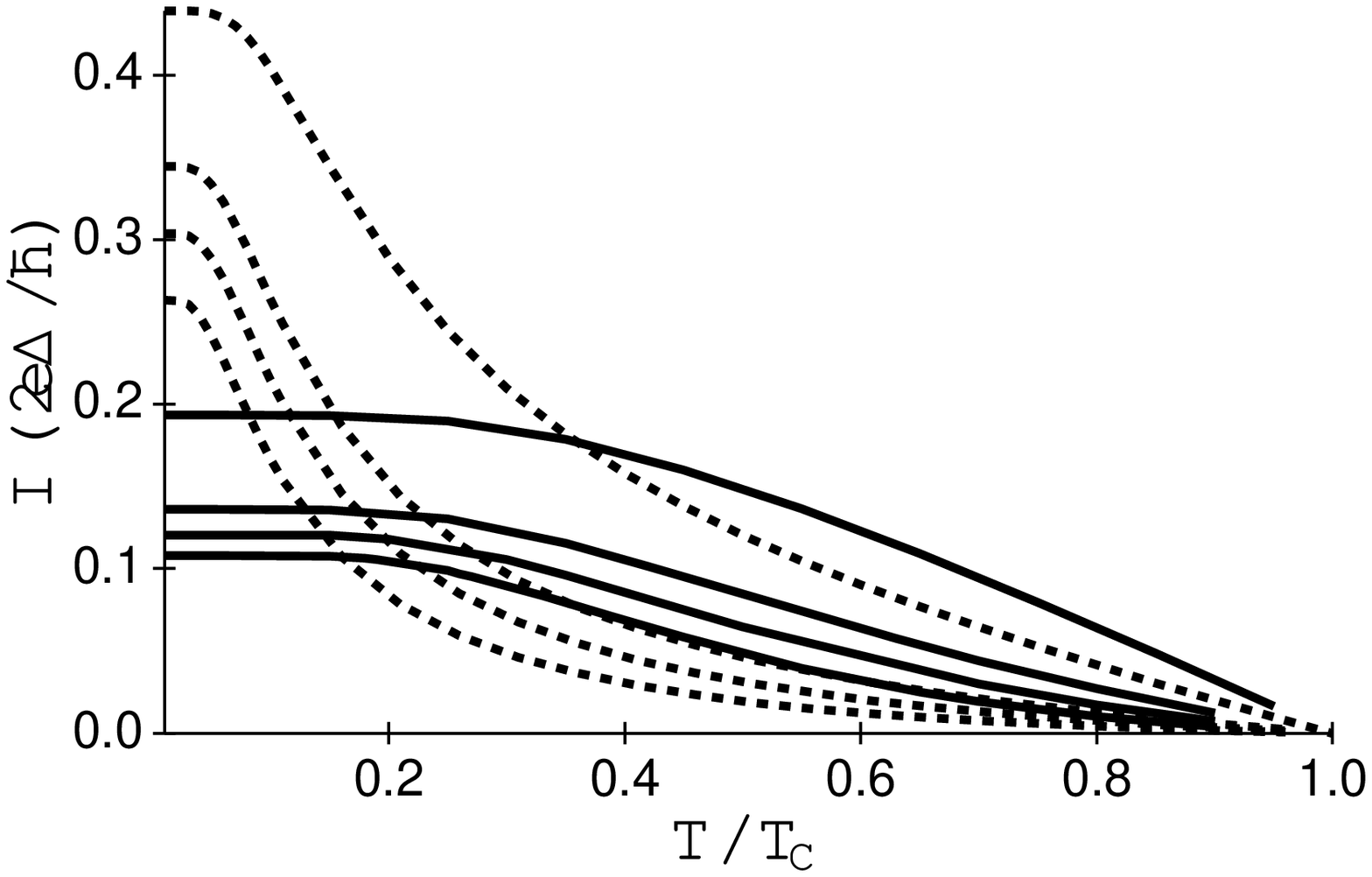}
\end{center}
\begin{small} {\bf Fig. 5} Temperature dependence 
of the maximum Josephson current
for s-wave (dashed curves) and p-wave(solid curves) cases
when $t_{K}=$100, 5, 3 and 2 (from the top to the bottom).
The phase $\delta$ of each curve is chosen so as to
give the maximum value of Josephson current at $T=0$ in Fig.~4.
The rest parameters are the same as in Fig.~1. \\
\end{small}
\vskip -0.5 truecm
By the same
reasoning the temperature dependence (see Fig. 5) of 
the Josephson current is more
similar to the 
usual SIS junction for the $p-wave$ case, while for $s-wave$
superconductors the current behavior is closer to the temperature
dependence of the Josephson Current in SNS junctions.
To be more precise, Fig. 5 displays the
temperature dependence of 
the maximum Josephson current $I(T)$ 
for s-wave (dashed lines) and p-wave (solid lines) cases.
For p-wave junctions,
$I(T)$ assumes relatively large value near $T=T_C$
while for s-wave junctions, it rapidly deceases as $T$ increases.
This difference stems from the question of
whether the mid-gap state appears or not.
The situation is similar to the one encountered in
Josephson current through in an SIS system
~\cite{tanaka&kashiwaya}.
The difference becomes more prominent when
the Kondo effect is suppressed;
$I(T)$ drops more rapidly for s-wave junctions than for p-wave ones
as $t_K$ decreases.

In conclusion, we have analyzed an important physical problem
involving strong correlations, the Kondo effect and
superconductivity. These aspects can be combined in an $SKN$
junction consisting of an Anderson impurity (in the Kondo regime).
We have developed a theoretical framework by which it is possible
to investigate an interplay between Andreev reflections and the
formation of Kondo resonance in the Kondo regime $T < \Delta <
T_{K}$. In this limit we calculated Josephson Current. We have also
investigated non-equilibrium aspects and elucidated
the nonlinear $G-V$ characteristics. Moreover, we calculated
the shot-noise power spectrum of $SKN$ junctions at voltages $eV
\leq \Delta$. It is found that at sufficiently large $t_K$ the
Kondo resonance effectively turns the junction behavior 
to be similar to
that of highly transparent non-interacting weak links for both $s$
and $p-wave$ superconductors. However, when the ratio of the Kondo
temperature to the superconducting gap becomes smaller the
behavior of these two types of junctions is quite different: the
Kondo resonance persists much more effectively for 
junctions with $p-wave$ leads than it does for $s-wave$ leads.

\noindent 
{\bf Acknowledgment} We would like to thank Y. Tanaka for 
very helpful and stimulating discussions. 
 This research is supported by 
DIP German Israel Cooperation project {\bf Quantum electronics in
low dimensions} by the Israeli Science Foundation grant {\bf
Many-Body effects in non-linear tunneling} and by the US-Israel
BSF grant {\bf Dynamical instabilities in quantum dots}. 
One of us (TA) is supported by
JSPS Research Fellowships for Young Scientists.

\end{document}